\documentclass[final,5p,times,twocolumn]{elsarticle}
	\usepackage{color}
	\usepackage{arydshln}
	\usepackage{lineno}

    \journal{}
    
	\newcommand{\terrormax}{1.5}
	\newcommand{\etal}{{\it et al.}}

\begin{document}
	
	\begin{frontmatter}
	
	\title{A measurement of the time profile of scintillation induced by low energy gamma-rays in liquid xenon with the XMASS-I detector}

	\address{\rm\normalsize XMASS Collaboration$^*$}
	\cortext[cor1]{{\it E-mail address:} xmass.publications3@km.icrr.u-tokyo.ac.jp .} 
	
	\author[ISEE]{H.~Takiya}
	\author[ICRR,IPMU]{K.~Abe}
	\author[ICRR,IPMU]{K.~Hiraide}
	\author[ICRR,IPMU]{K.~Ichimura}
	\author[ICRR,IPMU]{Y.~Kishimoto}
	\author[ICRR,IPMU]{K.~Kobayashi}
	\author[ICRR]{M.~Kobayashi}
	\author[ICRR,IPMU]{S.~Moriyama}
	\author[ICRR,IPMU]{M.~Nakahata}
	\author[ICRR]{T.~Norita}
	\author[ICRR,IPMU]{H.~Ogawa}
	\author[ICRR,IPMU]{H.~Sekiya}
	\author[ICRR]{O.~Takachio}
	\author[ICRR,IPMU]{A.~Takeda}
	\author[ICRR]{S.~Tasaka}
	\author[ICRR,IPMU]{M.~Yamashita}
	\author[ICRR,IPMU]{B.~S.~Yang}
	\author[IBS]{N.~Y.~Kim}
	\author[IBS]{Y.~D.~Kim}
	\author[ISEE,KMI]{Y.~Itow}
	\author[ISEE]{R.~Kegasa}
	\author[ISEE]{K.~Kobayashi}
	\author[ISEE]{K.~Masuda}
	\author[Tokushima]{K.~Fushimi}
	\author[IPMU]{K.~Martens}
	\author[IPMU]{Y.~Suzuki}
	\author[Kobe]{R.~Fujita}
	\author[Kobe]{K.~Hosokawa}
	\author[Kobe]{K.~Miuchi}
	\author[Kobe]{N.~Oka}
	\author[Kobe]{Y.~Onishi}
	\author[Kobe,IPMU]{Y.~Takeuchi}
	\author[KRISS,IBS]{Y.~H.~Kim}
	\author[KRISS]{J.~S.~Lee}
	\author[KRISS]{K.~B.~Lee}
	\author[KRISS]{M.~K.~Lee}
	\author[Miyagi]{Y.~Fukuda}
	\author[Tokai1]{K.~Nishijima}
	\author[YNU1]{S.~Nakamura}

	\address[ICRR]{Kamioka Observatory, Institute for Cosmic Ray Research, the University of Tokyo, Higashi-Mozumi, Kamioka, Hida, Gifu, 506-1205, Japan}
	\address[IBS]{Center of Underground Physics, Institute for Basic Science, 70 Yuseong-daero 1689-gil, Yuseong-gu, Daejeon, 305-811, South Korea}
	\address[ISEE]{Institute for Space-Earth Environmental Research, Nagoya University, Nagoya, Aichi 464-8601, Japan}
	\address[Tokushima]{Institute of Socio-Arts and Sciences, The University of Tokushima, 1-1 Minamijosanjimacho Tokushima city, Tokushima, 770-8502, Japan}
	\address[IPMU]{Kavli Institute for the Physics and Mathematics of the Universe (WPI), the University of Tokyo, Kashiwa, Chiba, 277-8582, Japan}
	\address[KMI]{Kobayashi-Maskawa Institute for the Origin of Particles and the Universe, Nagoya University, Furo-cho, Chikusa-ku, Nagoya, Aichi, 464-8602, Japan}
	\address[Kobe]{Department of Physics, Kobe University, Kobe, Hyogo 657-8501, Japan}
	\address[KRISS]{Korea Research Institute of Standards and Science, Daejeon 305-340, South Korea}
	\address[Miyagi]{Department of Physics, Miyagi University of Education, Sendai, Miyagi 980-0845, Japan}
	\address[Tokai1]{Department of Physics, Tokai University, Hiratsuka, Kanagawa 259-1292, Japan}
	\address[YNU1]{Department of Physics, Faculty of Engineering, Yokohama National University, Yokohama, Kanagawa 240-8501, Japan}

	\begin{abstract}
	We report the measurement of the emission time profile of scintillation from gamma-ray induced events in the XMASS-I 832~kg liquid xenon scintillation detector.
	Decay time constant was derived from a comparison of scintillation photon timing distributions between the observed data and simulated samples in order to take into account optical processes such as absorption and scattering in liquid xenon.
	Calibration data of radioactive sources, $^{55}$Fe, $^{241}$Am, and $^{57}$Co were used to obtain the decay time constant.
	Assuming two decay components, $\tau_1$ and $\tau_2$, the decay time constant $\tau_2$ increased from
	27.9~ns to 37.0~ns as the gamma-ray energy increased from 5.9~keV to 122~keV.
	The accuracy of the measurement was better than \terrormax~ns at all energy levels.
	A fast decay component with $\tau_1 \sim 2$~ns was necessary to reproduce data.
	Energy dependencies of $\tau_2$ and the fraction of the fast decay component were studied as a function of
	the kinetic energy of electrons induced by gamma-rays.
	The obtained data almost reproduced previously reported results and extended them to the lower energy region
	relevant to direct dark matter searches.
	\end{abstract}
	

	\end{frontmatter}

    
	\section{Introduction}
	Liquid xenon (LXe) has been used in many experiments for dark matter searches \cite{bibXMASSDetector, bibLUXDetector, bibXENON100Detector, bibPandaXDetector}, double beta decay searches \cite{bibEXODetector} and lepton flavor violation searches \cite{bibMEGDetector}.
	The time profile of LXe scintillation is important information for these experiments.
	It could potentially be used for particle identification~\cite{bibPSD} and vertex reconstruction~\cite{bibCLEAN, bibInelasticWIMP}.
	
	Basic characteristics of scintillation emission in LXe have been intensively studied elsewhere in order to understand the detector response.
	There are two scintillation processes in LXe, the direct scintillation and the recombination processes.
	The direct scintillation process proceeds through two states, singlet excitation $^1\Sigma^+_{\mu}$ and triplet excitation $^3\Sigma^+_{\mu}$.
	The decay time constants of singlet and triplet states are a few ns and $\sim$20~ns, respectively \cite{bibKubota, bibHitachi}.
	The recombination process has a longer decay time constant of $\sim$30~ns or more \cite{bibKubota, bibHitachi}.
    The scintillation time profile can be used to discriminate between nuclear recoil events and electron events since
    the ratio of singlet to triplet excitations as well as the recombination time depend on ionization density~\cite{bibHitachi,bibKubota2}. 

	Existing time profile measurements have been conducted with a small amount of
	LXe~\cite{bibKubota,bibHitachi,bibKeto,bibAkimov,bibDawson,bibUeshima,bibTeymourian,bibMurayama}.
	This method minimizes the scattering and absorption of scintillation photons by xenon itself.
	Some prior research, however, reported that decay time constants do not agree with each other,
	as reviewed in~\cite{bibMurayama, bibNEST}.
	The disagreements might be caused by differences in the experimental setups, conditions of the LXe, or analysis methods.
	Furthermore, the fast decay component and the energy dependence of the decay time constant must be considered \cite{bibKubota, bibAkimov}.
    Moreover, most of the previous measurements were performed with relatively small photoelectron yield
    making it difficult to measure events induced by low energy particles.
	Therefore, detailed measurements with larger photoelectron yield are necessary.
	
	A measurement of the time profile of scintillation in LXe using the XMASS-I detector was conducted.
	The XMASS experiment is for direct dark matter search using the 832~kg of LXe scintillator \cite{bibXMASSDetector}.
	The XMASS-I detector has a large photo-coverage of more than 62$\%$.
	Owing to the large amount of LXe and large photo-coverage, it is possible to obtain the time profile measurements.
	In this paper, radioactive sources, $^{55}$Fe, $^{241}$Am and $^{57}$Co, were used to measure the time profile for a wide energy range, between 5.9~keV and 122~keV as gamma-ray energy.

	\section{The XMASS experiment}		
	The XMASS detector consists of a copper vessel surrounded by a large water tank for shielding \cite{bibXMASSDetector}.
	The LXe in the inner vessel is viewed by 642 photomultiplier tubes (PMTs).
	The PMTs are implemented into PMT holders made of oxygen free high conductivity copper.
	The holders are assembled into a pentakis dodecahedron surrounding LXe
	for maximizing the collection efficiency of the scintillation photons.
	Xenon was purified by two getters (PS4-MT15, SAES) before filling into the detector.
	As a result, the light yield is quite high, approximately 14 photoelectrons (PE)/keV.
	
	A radioactive source can be inserted into the detector for the purpose of calibration.
	The source position is movable only along the $Z$ (vertical) axis.	
	The detector center is at $Z=0$~cm.
	The sources can be divided into two groups according to their structure.
	All of them are mounted in the needle-shaped containers with different diameters.
	The 2$\pi$ sources, $^{55}$Fe and $^{241}$Am (2$\pi$), have a 10~mm diameter.
	The 4$\pi$ sources, $^{241}$Am (4$\pi$) and $^{57}$Co, have a 0.21~mm diameter \cite{bibXMASSDetector, bibXMASSMicroSource}.	
	The two types of sources were developed to better handle the shadow effect from the source itself.
	A thin source structure is preferred because it is better at avoiding the shadow effect by source itself.
	However, in the case of low energy radiation, interactions occur close to the source due to the short attenuation length.
	Therefore, the shadow effect can be observed even for a thin structure.
	The uncertainties caused by roughness of the source surface must be considered.
	While it is difficult to polish a thin structure, $2\pi$ sources make handling the uncertainties easier due to their well polished flat surfaces.
	$^{55}$Fe decays into $^{55}$Mn via electron capture and 5.9~keV characteristic X-rays are emitted.
	$^{241}$Am decays into its daughter nuclei $^{237}$Np.
	$^{237}$Np emits 59.5~keV gamma-rays and 17.8~keV X-rays.
	While both of the 59.5~keV gamma-rays and 17.8~keV X-rays are observable in the case of $^{241}$Am ($4\pi$), 17.8~keV X-rays are not observable in the case of $^{241}$Am ($2\pi$) due to the thick structure.
	When a 59.5~keV gamma-ray is absorbed in LXe, a 25.0~keV electron is emitted from the $K$-shell due to the photoelectric effect,
	and an approximately 30~keV characteristic X-ray and low energy Auger electrons are emitted.
	In the case of the $^{241}$Am ($2\pi$) source, the X-rays often escape from LXe back into the source itself
	due to the large solid angle of the source, and therefore an ``escape peak'' can be observed at the deposited energy of $\sim$30~keV.	
	$^{57}$Co emits 122~keV gamma-rays and 59.3~keV X-rays from tungsten contained in the source.
	
	The signals from the PMTs pass through $\sim$20~m coaxial cables to CAEN V1751 waveform digitizers.
	The waveforms in each PMT are recorded with 1~GHz sampling rate and 10~bit resolution.
	The threshold for a PMT is set to $-5$~mV and it corresponds to 0.2~PE.
	A trigger is issued when at least four PMTs detect signals exceeding the threshold within 200~ns.
	A detailed explanation is provided in \cite{bibXMASSDetector}.
	
	Timing calibration with $^{57}$Co is regularly carried out to adjust the timing offset of each PMT channel due to the differences in their cable lengths (at most 2~m) and responses of the electronics.
	The $^{57}$Co source is placed at $Z=0$~cm where the distance to each PMT is nearly equal.
	With approximately 10,000 events in the 122~keV gamma-ray peak,
	the distribution of the threshold crossing time for each channel is fitted with a combination of two exponential functions convoluted with a Gaussian
	to get the timing offset of each channel so that the rising edges of the distributions are aligned.
	The precision of the calibration is better than 0.3~ns, estimated from the uncertainty in the fitting.
	PMT gain stability is monitored using signals generated by a blue LED implemented on the inner surface of the detector.

	\section{Analysis method}

    \begin{figure}[tbp]
		\begin{center}
			\includegraphics[width=95mm]{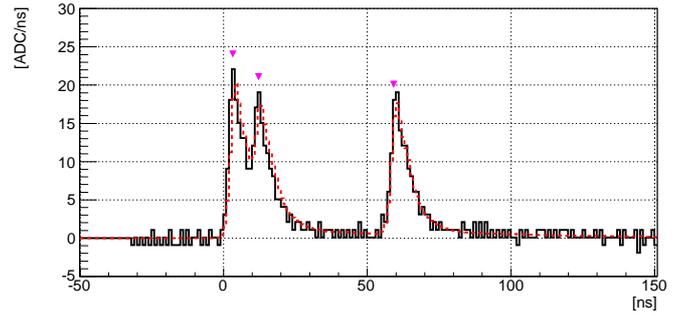}	
		\end{center}
		\caption{A typical raw waveform in a PMT (solid line) overlaid with the reconstructed waveform
		as the sum of 1~PE pulses (dashed curve) for the 122~keV gamma-ray from the $^{57}$Co source placed at $Z$=0~cm.
		The triangle markers indicate timings of the decoupled 1~PE pulses.}
		\label{fig:Waveform}
	\end{figure}		
	
	The scintillation time profile is evaluated by comparing the reconstructed pulse timing distributions over all PMTs of data and simulated samples with various timing parameters.
	Pulse splitting method has been developed in the XMASS experiment which enabled the ability to obtain peak timing of each scintillation photon pulse.
	
	Pulse splitting is executed using a peak search algorithm based on Savitzky-Golay filter \cite{bibPulseSplitter}.
	The waveform data are fitted with a convolution of 1~PE pulse waveform obtained from LED calibration data.
	Waveform fitting using the 1~PE waveform template is done on a 1~ns grid.	
	Figure~\ref{fig:Waveform} shows a typical raw waveform in a PMT overlaid with the reconstructed waveform
	as the sum of the 1~PE pulses for the 122~keV gamma-ray from the $^{57}$Co source placed at $Z$=0~cm.
	It corresponds to 3~PE incident and the observed waveform can be clearly reconstructed as the sum of the three 1~PE pulses.
	Owing to a small fluctuation of the baseline, pulse splitting sometimes makes small artifact pulses in the data in the case of a large number of incident photons.
	These pulses clearly appear more than 60~ns after the primary and affect the apparent decay time constant.
	These pulses can be rejected on the basis that the PE of a pulse is larger than 0.5~PE.
	
	The XMASS Monte Carlo simulation is based on Geant4 \cite{bibGeant4}.
	The energy-dependent scintillation photon yield is taken into account using a non-linearity model from Doke \etal~\cite{bibDoke}
	with a further correction obtained from gamma-ray calibrations in the relevant energy range.
	A precise understanding of the optical characteristics inside the detector is needed to extract a time profile.
	Optical parameters of LXe and the inner surface material of the detector are also carefully tuned by source calibration data at various positions.

	A scintillation photon observed time $T$ in the simulation is defined as below.
	\begin{equation}
		T = t_{\rm{Edep}} + t_{\rm{scinti}} + t_{\rm{TOF}} + t_{\rm{TT}} + t_{\rm{elec}}
	\end{equation}
	Here $t_{\rm{Edep}}$ is time of energy deposition from the incident particle to xenon.
	$t_{\rm{scinti}}$ is a value that follows the scintillation time profile as explained later.
	$t_{\rm{TOF}}$ is the time of flight of the scintillation photons to each PMT.
	The group velocity of the scintillation light in LXe is calculated from its refractive index ($\sim$11~cm/ns for 175~nm wavelength \cite{bibFujii}).
	$t_{\rm{TT}}$ is the transit time in a PMT.
	It is a value that follows the transit time spread.
	It is randomly generated from the transit time spread distribution derived from simulation of the PMT done by manufacturer.
	$t_{\rm{elec}}$ is a timing smearing parameter to reproduce the uncertainty of the electronics simulation.
	$t_{\rm{elec}}$ is assumed to follow a Gaussian distribution with a standard deviation $t_{\rm{jitter}} = 0.93$~ns,
	evaluated from $^{57}$Co calibration data.
	The uncertainty of the electronics simulation is taken into account as a systematic error.
	Waveforms in each PMT are simulated taking into account the 1~PE pulse shape.
    After the simulation, the pulse splitting method is also applied to simulated samples.
	Additional details are provided in \cite{bibXMASSDetector}.
	
	The simulated samples with two decay components, $\tau_{1}$ and $\tau_{2}$, are used for this analysis.
	The fast decay component $\tau_{1}$ corresponds to the singlet excitation process $^1\Sigma^+_{\mu}$.
	$\tau_{1}$ values in prior research do not agree with each other.
	Kubota \etal~\cite{bibKubota} reported $\tau_{1} = 2.2\pm0.3$~ns with a $^{207}$Bi source and
	Hitachi \etal~\cite{bibHitachi} reported $\tau_{1} = 4.3\pm0.6$~ns with a $^{252}$Cf source.
	Here, $\tau_{1} = 2.2$~ns is chosen because only Kubota \etal \ reported fast decay component with electrons.
	The results, however, do not significantly change with the alternative assumption of $\tau_1 = 4.3$~ns.
	The slow decay component $\tau_{2}$ corresponds to the convolution of triplet excitation $^3\Sigma^+_{\mu}$ and recombination processes.
	
	Scintillation time $t_{\rm{scinti}}$ is described as equation (\ref{eq:fraction}).
	\begin{equation}
		f\left( t \right)	=
						\frac{F_1}{\tau_1}
							\exp \left( - \frac{t}{\tau_1} \right)
							+ \left( \frac{1 - F_1}{\tau_2} \right) \cdot
							\exp \left( - \frac{t}{\tau_2} \right)
		\label{eq:fraction}
	\end{equation}
	$F_1$ defines the fraction of photons following the fast decay time constant $\tau_{1}$ in an event.
	Namely, there are two free parameters, $\tau_{2}$ and $F_{1}$.
	$\tau_{2}$ and $F_{1}$ are scanned in the simulation with steps of 1~ns and 0.025, respectively.
	$\tau_{2}$ is changed from 20~ns to 30~ns, 25~ns to 35~ns, 30~ns to 40~ns for $^{55}$Fe, $^{241}$Am (2$\pi$ / 4$\pi$) and $^{57}$Co, respectively.
	$F_{1}$ is changed from 0.0 to 0.15 for each $\tau_{2}$ value.
	
	There are some other Monte Carlo parameters, scattering and absorption length, which may affect the time profile.
	Both are also considered in simulation.
	Using the $^{57}$Co calibration data of various positions,
	these parameters are tuned such that the observed number of PE in each PMT in the simulated samples reproduce those in data.
	The scattering length is almost stable at 52~cm within 1~cm, however, the absorption length varies from 4~m to 11~m.
	These parameters are independent from timing information.
	The uncertainties of the Monte Carlo parameters are taken into account as systematic errors.
	
	The reconstructed 1PE timing distributions in data and simulation are both made by summing up
	the reconstructed 1PE timing distributions from all PMTs. The reconstructed 1PE timing distribution
	from each PMT is corrected for the timing offset.	
	The fourth earliest pulse timing is adjusted to $T=0$~ns in every event to reflect the trigger implementation. 
	Agreement of the timing distributions is evaluated by $\chi ^2$ defined as below.
	\begin{equation}
		\chi ^2 = \sum_{i}
			\frac{( N^{\rm{data}}_i - N^{\rm{MC}}_i \times S ) ^2}
						{ N^{\rm{data}}_i + N^{\rm{MC}}_i \times S^2}
	\end{equation}
	Here $N^{\rm data}_i$ and $N^{\rm MC}_i$ are the number of pulses in the $i$-th time bin for data and Monte Carlo simulation, respectively.
	One bin corresponds to 1~ns, and $\chi^2$ is calculated for the bins corresponding to time range of 3~ns $\leq T \leq 120$~ns.
	The region is set to exclude a part of the rise edge that fluctuates and worsens $\chi^2$ largely.
	$S$ is a normalization factor defined as the ratio of the total number of 1~PE pulses in the data to that in the simulated samples,
	which is typically between 0.3 and 0.4 since the simulated samples contain an approximately three times larger number of events than data.
	
	The analyzed data were occasionally acquired from the end of 2013 to the beginning of 2015.
	It was after the refurbishment of the detector that the background was further reduced.
	Temperature and pressure were stable during this period, 172.6~K to 173.0~K and 0.162~MPa to 0.164~MPa, respectively.

	Event selections for this analysis are primarily done by the number of PE.
	We select events whose number of PE is within $\pm10$~PE from the peak in order to restrict the incident particle energy.
	Two more event selections are applied to reject events caused by the afterpulses of bright events.  
	The events occurred within 10~ms from the previous events, and the events whose root mean square of the timing distribution is larger than 100~ns, are rejected.
	As mentioned above, pulses whose number of PE are less than 0.5 are rejected.

	\section{Results and discussions}
\begin{figure}[tbp]
		\begin{center}
			\includegraphics[width=95mm]{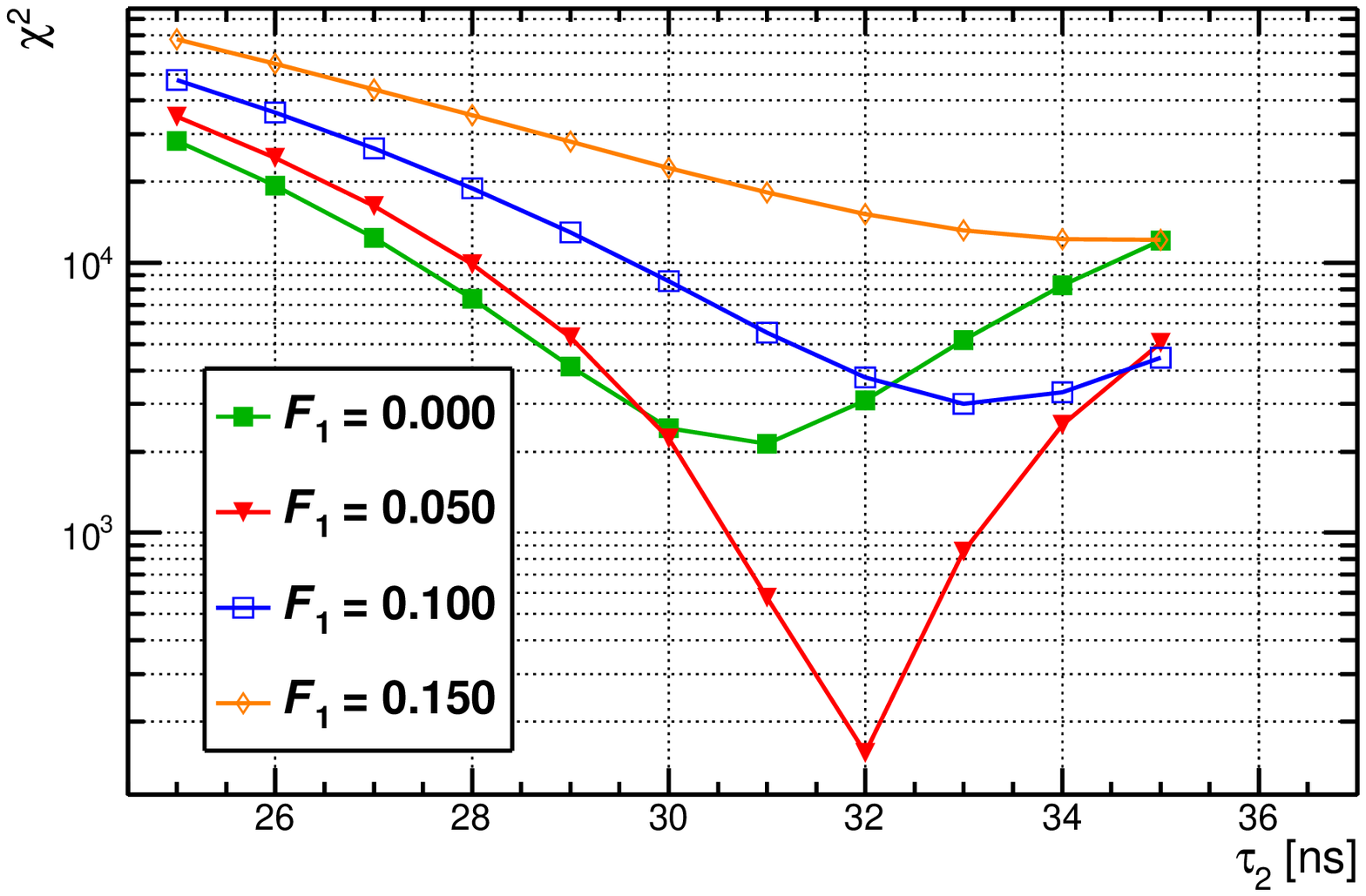}
			\includegraphics[width=95mm]{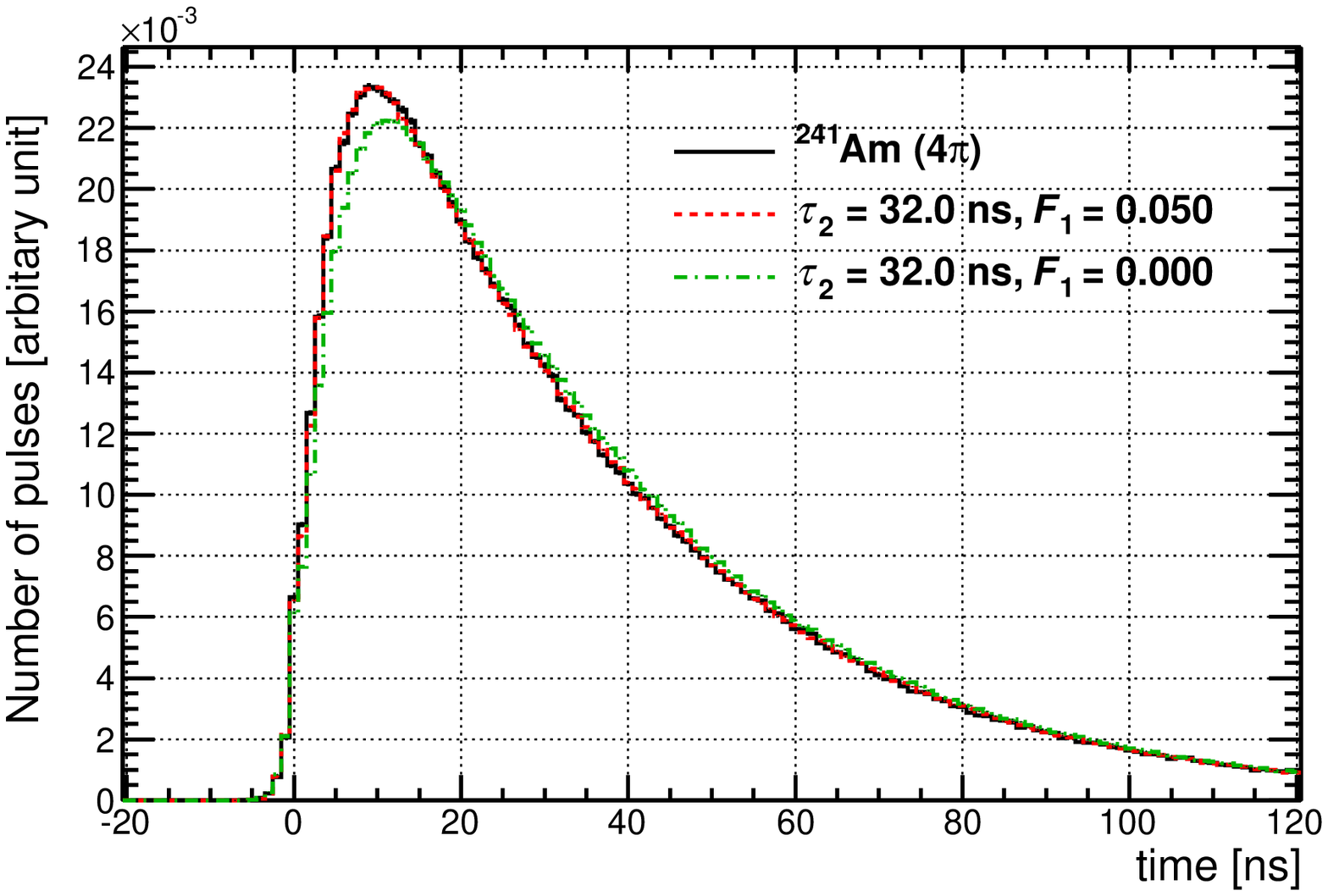}
			\includegraphics[width=95mm]{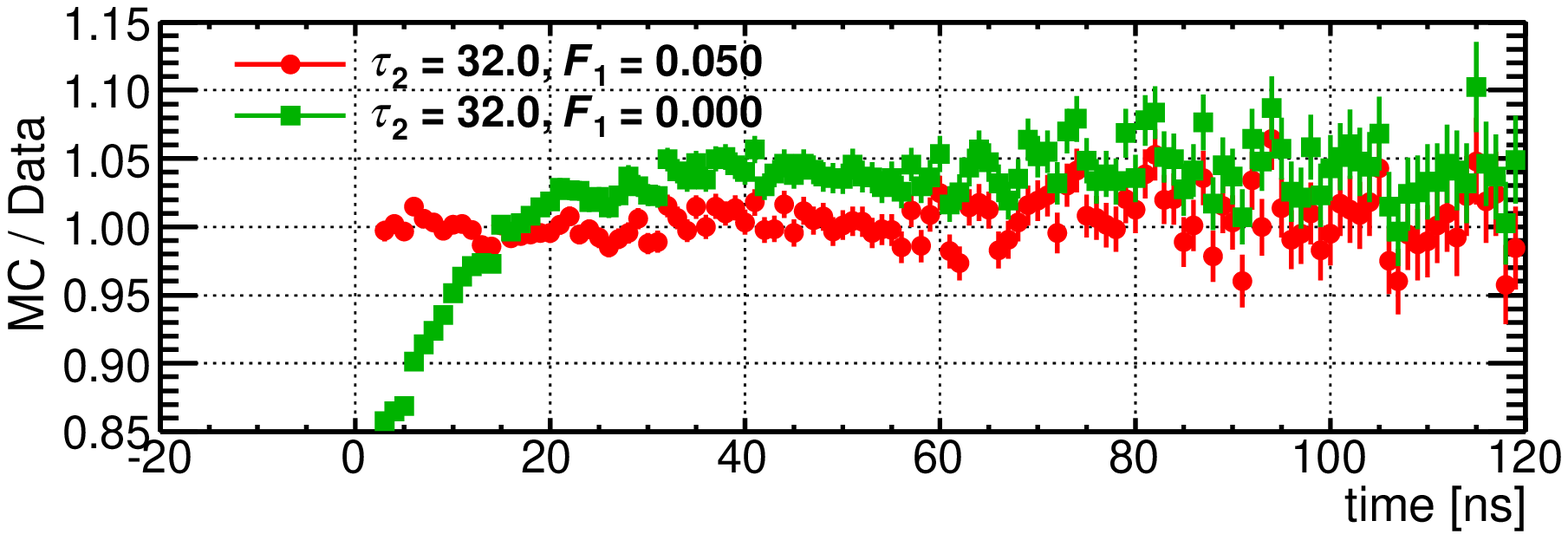}
		\end{center}
		\caption{(Top) $\chi ^2$ as a function of the decay time constant $\tau_2$ for various values of the fast decay component fraction $F_1$.
					 (Middle) The pulse timing distribution of the observed data overlaid with those of the best fit simulated samples with and without a fast decay component.
					 The fourth earliest pulse timing is shifted to $T=0$ ns for event-by-event timing adjustment.
					 (Bottom) Monte Carlo simulation over a data ratio of the pulse timing distributions.
					 A less than 3~ns region is not used for the $\chi^2$ calculation, therefore no points are drawn at the region.}
		\label{fig:241AmZ=00cm60keVResult}
	\end{figure}

	Figure \ref{fig:241AmZ=00cm60keVResult}~(top) shows $\chi^2$ as a function of $\tau_2$.
	The analyzed data are from the $^{241}$Am (4$\pi$) source placed at the center of the detector, $Z=0$~cm.
	The induced gamma-ray energy $E_{\gamma}=59.5$~keV, corresponds to 750 to 770~PE.
	A simulated sample with $\tau_2 = 32$~ns and $F_1 = 0.05$ gives the minimum $\chi^2/\rm{dof} = 153.0 / 118$.
	The pulse timing distribution of the observed data overlaid with those of the best fit simulated
	samples with and without a fast decay component is shown in Figure~\ref{fig:241AmZ=00cm60keVResult}~(middle).
	Simulated samples without a fast decay component gives $\chi^2 = 3094.6$ at the best with $\tau_2 = 32$~ns due to the discrepancy for $t < 15$~ns.
	Therefore, a fast decay component is necessary to reproduce data. 	
	The more precise $\tau_2$ and $F_1$ values in between the two steps are estimated by interpolating $\chi^2$ values with a one-dimensional quadratic function, resulting in $\tau_2=31.9$~ns and $F_1=0.048$.

	All the systematic errors are itemized in Table \ref{tbl:ErrorSummary}.
	The position dependence is less than 1.3~ns and 0.017 for $\tau_2$ and $F_1$, respectively.
	It is evaluated from the standard deviation of the measured values in $^{241}$Am (4$\pi$) data.
	The position dependence evaluated from 17.8~keV gamma-ray is applied for the gamma-ray energies of $E_{\gamma} = 17.8$~keV or less ($E_{\gamma} = 5.9$, 17.8~keV).
	On the other hand, position dependence evaluated from 59.5~keV gamma-ray is applied for mean kinetic electron energies of $E_{\rm{electron}} \sim22$~keV or more ($E_{\gamma} = 59.5$, 59.3, 122~keV and escape electron $E_{\rm{electron}} \sim 22$~keV from $^{241}$Am (2$\pi$)).
	Possible differences due to the source shape are evaluated from $Z=0$~cm data for $^{241}$Am ($4\pi$ / $2\pi$) and $^{57}$Co.
	The uncertainty is $\pm0.9$~ns and $\pm0.004$ for $\tau_2$ and $F_1$, respectively.
	Timing jitter $t_{\rm{jitter}}$ mainly affects the rising edge of the timing distributions.
	The error is evaluated by comparing the timing distributions of data and simulated samples with different assumptions for $t_{\rm{jitter}}$.
	$t_{\rm{jitter}}$ is changed to 0.0, 0.5, and 1.5~ns.
	The systematic error is less than 0.2~ns for $\tau_2$, and less than 0.009 for $F_1$.
	The errors caused by the optical parameters in the Monte Carlo simulation are also evaluated as well as timing jitter.
	The scattering length is changed by $\pm1$~cm.
	The absorption length is changed from 6~m to 4~ m and 11~m.
	As a result, the errors caused by the optical parameters are less than 0.2~ns and 0.014 for $\tau_2$ and $F_1$, respectively. 
	The detector stability including light yield change is evaluated by applying the same analysis method to other calibration data taken at different times.
	Nine other data sets of $^{57}$Co are used for the evaluation.
	The errors are found to be less than 0.9~ns and less than 0.009 for $\tau_2$ and $F_1$, respectively.
	The errors caused by step size of $\tau_2$ and $F_1$ values, 1.0~ns and 0.025, are evaluated from the differences between the parameters of the simulated samples
	which gives minimum $\chi^2$ and the values obtained by interpolation.
	The errors are $\pm0.3$~ns for $\tau_2$, and $\pm0.007$ for $F_1$.

	\begin{table*}[tbp]
		\begin{center}
			\caption{Summary of systematic uncertainties on $\tau_2$ and $F_1$.}
			\begin{tabular}{lccc}
				\hline \hline
					\\[-3.5mm]
					Error source	&	$\sigma_{\tau_2}$	(ns)	&	$\sigma_{F_1}$	&	Energy range	\\
				\hline \\[-3mm]
					Position dependence
							&	$\pm0.9$	&	$^{+0.017}_{-0.011}$	
								&	$E_{\gamma} \leq 17.8$ keV	\vspace{1.2mm}	\\
							&	$^{+0.3}_{-1.3}	$	&	$^{+0.005}_{-0.004}$	
								&	$E_{\gamma} \geq 59.3$ keV	\vspace{1.2mm}	\\
					\hdashline	\\[-3mm]
					Source difference
							&	$\pm0.6$	&	$\pm0.002$	&	All	\vspace{1.2mm}	\\
					\hdashline	\\[-3mm]
					Timing jitter
							&	$\pm0.1$	&	$\pm0.009$	&	$E_{\gamma} \leq 17.8$ keV	\vspace{1.2mm}\\
							&	$\pm0.2$	&	$\pm0.001$	&	$E_{\gamma} \geq 59.3$ keV	\vspace{1.2mm}\\
					\hdashline	\\[-3mm]
					Optical parameters
							&	$\pm0.2$	&	$\pm0.007$	&	$E_{\gamma} \leq 17.8$ keV	\vspace{1.2mm}	\\
							&	$\pm0.1$	&	$\pm0.014$	&	$E_{\gamma} \geq 59.3$ keV	\vspace{1.2mm}	\\
					\hdashline	\\[-3mm]
					Detector stability
							&	$^{+0.9}_{-0.0}$	&	$^{+0.005}_{-0.009}$	
								&	$E_{\gamma} \leq 59.5$ keV	\vspace{1.2mm}	\\
							&	$^{+0.4}_{-0.0}$	&	---	
								&	$E_{\gamma} = 122$ keV	\vspace{1.2mm}	\\
					\hdashline \\[-3mm]
					Step size
							&	$\pm0.3$	&	$\pm0.007$	&	All	\vspace{1.2mm}	\\
					\hline \\[-3.3mm]
							&	$^{+1.5}_{-1.1}$	&	$^{+0.022}_{-0.020}$	
								&	$E_{\gamma} \leq 17.8$ keV	\vspace{1.2mm}\\
					Total	&	$^{+1.2}_{-1.5}$	&	$^{+0.017}_{-0.019}$
								&	$E_{\gamma} = 59.3, 59.5$ keV	\vspace{1.2mm}\\
							&	$^{+0.9}_{-1.5}$	&	---	
								&	$E_{\gamma} = 122$ keV	\vspace{1.2mm}	\\
				\hline \hline
			\end{tabular}
			\label{tbl:ErrorSummary}
		\end{center}
	\end{table*}

	The measured decay time constant values are summarized in Table \ref{tbl:DecayConstantSummary}.
	\begin{table*}[tbp]
		\begin{center}
			\caption{Summary of the decay time constant and incident particle energy.
						Incident electron energy $E_{\rm{electron}}$ is evaluated by Monte Carlo simulation.
						See text for the uncertainty in $E_{\rm{electron}}$.
						No $F_1$ value is shown in $E_{\gamma} = 122$ keV of $^{57}$Co because
						the time range for the $\chi^2$ calculation is changed.}
			\begin{tabular}{lccccc}
				\hline \hline
					\\[-2.5mm]
					Source	&	$E_{\gamma}$ (keV)	&	$E_{\rm{electron}}$ (keV)	&	$\tau_{2}$ (ns)	& $F_1$	&	\\
					\\[-3.5mm]
				\hline
					\\[-3mm]
					$^{55}$Fe	&	5.9	&	$3.3\pm1.3$	&	$27.8^{+1.5}_{-1.1}$	
												&	$0.145^{+0.022}_{-0.020}$	&	\vspace{1.2mm}\\
					$^{241}$Am	&	17.8	&	$12.2\pm4.6$	&	$27.9^{+1.5}_{-1.1}$	
												&	$0.098^{+0.022}_{-0.020}$	&	\vspace{1.2mm}\\
									& ---		&	$22.0\pm7.2$	&	$32.2^{+1.2}_{-1.5}$	
												&	$0.063^{+0.017}_{-0.019}$	
												&	Escape electron from Xe	(2$\pi$ only)\vspace{1.2mm}\\
									&	59.5	&	$27.2\pm12.7$	&	$31.9^{+1.2}_{-1.5}$	
												&	$0.048^{+0.017}_{-0.019}$	&	\vspace{1.2mm}\\
					$^{57}$Co	&	59.3	&	$27.8\pm13.2$	&	$31.1^{+1.2}_{-1.5}$
												&	$0.045^{+0.017}_{-0.019}$	
												&	$K_{\rm{\alpha}}$ X-ray from tungsten	\vspace{1.5mm}\\
									&	122	&	$71.2\pm32.0$	&	$37.0^{+0.9}_{-1.5}$	
												&	---	&	$T \geq 30$ ns	\vspace{1.2mm}\\
				\hline \hline
			\end{tabular}
			\label{tbl:DecayConstantSummary}
		\end{center}
	\end{table*}
	A clear energy dependence on the decay time constant $\tau_2$ is found.
	Such energy dependence at the energy region has already been reported by Akimov \etal \cite{bibAkimov} and Ueshima \cite{bibUeshima}. 
	It would suggest that the fast decay component from the singlet state becomes visible because of a much shorter recombination time scale for the larger ionization density by a lower energy electron track.
	$F_1$ decreases as the incident particle energy increases.
	Note that the fitting is performed in the tail region of $T \geq 30$~ns for 122~keV because of a bad $\chi^2$ for fitting in the entire time range.
	Therefore, $F_1$ is not shown here.
	This might imply that a single exponential decay with $\tau_1$ may not be the case for
higher incident energy, possibly because the recombination process is more complex in this case.

	Energy dependencies of $\tau_2$ and $F_1$ are studied as a function of the kinetic energy of electrons induced by gamma-rays.
	In the case of 59.5~keV gamma-rays, 25.0~keV electrons are emitted from the $K$-shell, whose electron binding energy is 34.56~keV,
	due to the photoelectric effect.
	Auger electrons, whose energy are $\sim$25~keV or less are also emitted.
	Thus, multiple electrons with various kinetic energies can be emitted from an incident gamma-ray.
	The mean kinetic energy of the electrons $E_{\rm{electron}}$ is evaluated from a Monte Carlo simulation.
	The uncertainty for $E_{\rm{electron}}$ is defined as the root mean square of released electron energy.

	The energy dependence of the decay time constant in the $E_{\rm{electron}} < 100$~keV region has already been reported in prior research \cite{bibNEST, bibAkimov, bibUeshima}.
	The difference between $E_{\gamma}$ and $E_{\rm{electron}}$ for $\tau_2$ might be used for some experiments, such as specific dark matter searches \cite{bibInelasticWIMP, bibBosnicSuperWIMP}, and two neutrino double electron capture searches \cite{bibTwoNeutrinoDoubleElectronCapture}, by discriminating the gamma-ray induced events and electron induced events.
	
\begin{figure}[tbp]
		\begin{center}
			\includegraphics[width=90mm]{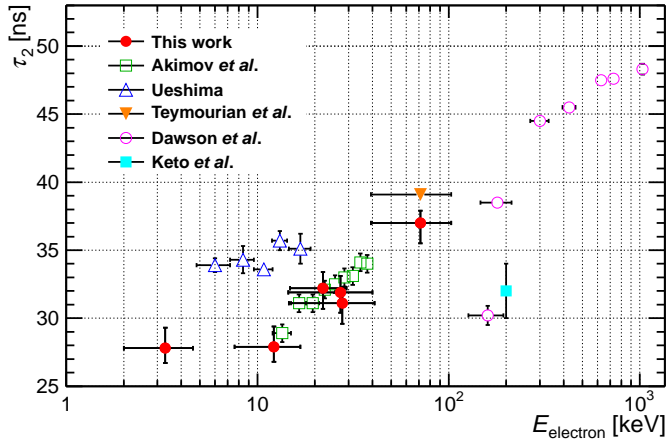}
		\end{center}
		\caption{$\tau_2$ as a function of incident electron energy $E_{\rm{electron}}$.
					Teymourian \etal \cite{bibTeymourian} used data from 122~keV gamma-ray from $^{57}$Co so the same error as this work was applied.
					Dawson \etal \cite{bibDawson} and Keto \etal \cite{bibKeto} reported $\tau_2$ at higher the energy region $E_{\rm{electron}} > 100$~keV.
					Some references \cite{bibKubota, bibHitachi, bibMurayama} are not drawn because $E_{\rm{electron}}$ is unknown.}
		\label{fig:ResultsWithPriorResearches}
	\end{figure}

	Figure~\ref{fig:ResultsWithPriorResearches} shows the decay time constant $\tau_2$ as a function of the electron kinetic energy $E_{\rm{electron}}$.
	This analysis gave consistent results with Akimov \etal~\cite{bibAkimov}
	and extended them to the lower energy region relevant to direct dark matter searches.
    Ueshima~\cite{bibUeshima} reported a much longer decay time constant than this analysis.
    It was found that this previous result was not corrected for the detector response:
    the observed waveform, without accounting for the detailed detector response such as the 1~PE waveform shape,
    was fitted by a single exponential function.
    The present analysis, on the other hand, decomposed waveforms into 1~PE pulses and compared their timing distributions
    between data and simulation to account for the detector response.
    
	\section{Conclusions}
	The time profile of scintillation in liquid xenon has been measured with the XMASS-I detector.
	The measurement was conducted in a wide energy range, between 5.9~keV to 122~keV for gamma-ray energy, with various radioactive sources, $^{55}$Fe, $^{241}$Am, and $^{57}$Co.
	Energy dependence of the decay time constant was observed.
	The decay time constant increased from 27.8~ns to 37.0~ns, and the error was smaller than \terrormax~ns.
	The obtained decay time constants are consistent with Akimov {\it et al}., but inconsistent with Ueshima.
	The discrepancy could be explained by the difference in the analysis methods.
	In addition, the 2.2~ns fast decay component, which corresponds to singlet excitation, was necessary to reproduce data.
	The number of photons that follow the fast decay component relatively decreased as incident particle energy increased.
	The ratio differed from 0.15 to 0.05 at the measured energy region.
	
	The measurements in this study provided a time profile of LXe scintillation below 10~keV with induced gamma-ray energy and revealed an energy dependence of $\tau_2$ and $F_1$.
	They are important for pulse shape discrimination of nuclear recoil from the gamma-ray signal, and also for possible discrimination of electron incident from gamma-ray incident, or vertex reconstruction using the scintillation time profile in the experiments such as dark matter and rare decay searches.

	\section*{Acknowledgements}
	We gratefully acknowledge the cooperation of Kamioka Mining  and  Smelting  Company.
   This work was supported by the Japanese Ministry of Education, Culture, Sports, Science and Technology, Grant-in-Aid for Scientific Research, JSPS KAKENHI Grant Number, 19GS0204, 26104004, and partially by the National Research Foundation of Korea Grant funded by the Korean Government (NRF-2011-220-C00006).


	\end{document}